# Evolution of clusters in energetic heavy ion bombarded amorphous graphite-III: Energy spectra of charged clusters recoiling from Xe⁺ irradiations

Bashir Ahmad[1], Maqsood Ahmad[1] and Shoaib Ahmad[2]

[1]PINSTECH, P. O. Nilore, Islamabad, Pakistan

[2]National Centre for Physics, Quaid-i-Azam University Campus, Shahdara Valley, Islamabad, 44000, Pakistan

Email: sahmad.ncp@gmail.com

## Abstract

We report on a novel technique of heavy ion induced surface modification and clustering mechanisms in amorphous graphite using Direct Recoil Spectrometry. The charged clusters emanating as Direct Recoils $-DRs$ from $50-150 keV\ Xe^+$ irradiated graphite surface are detected in successive energy spectra at large recoil angles. Ion beams irradiated surface is shown to provide an environment of sputtered carbon atoms and clusters. The clustering occurs on the surface where the most stable ones accumulate. The $DR$ spectra have identified clustering mechanisms operating in amorphous graphite under high fluence irradiations at grazing incidence that are related to the reconstructed surface features. Heavy ion irradiation induces the twin effects of producing a wide range of clusters and simultaneously ejecting these as DRs.

# I. Introduction

## A. Ion induced clustering phenomena in graphite and Carbon containing solids

Clustering of carbon atoms in polymers under energetic ion irradiation is a well documented phenomenon that can lead to optical blackening, electrical conductivity changes and has also been studied for ion induced chemical effects [1,2]. The mechanisms of nuclear as well as electronic energy transfer from ion to the carbon atoms in the solid matrix have been invoked to explain the ion-induced clustering processes. Orders of magnitude estimate for the size of graphitic islands or carbon rich zones range from 100-500 Å [2].

Recent experiments with MeV heavy ion sputtering of polymers at Uppsala [3,4] have indicated the formation of fullerenes in MeV Iodine ion bombarded PVDF targets. Positively charged even numbered carbon clusters $C_m^+$ ($m \geq 40$) have been detected in their time-of-flight spectra. The fullerene yield measurements as a function of ion fluence indicate the clustering to be dependent on ion induced chemical changes in the polymer. Chadderton et al [5] have reported the synthesis of fullerenes after 130 MeV/amu $Dy^{22+}$ ion bombardment of graphite. Chromatography of their irradiated samples has shown traces of $C_{60}$.

In our recent work at PINSTECH, using 100 keV Ar+, Kr+ and Xe+ beams on amorphous graphite, we have seen clear evidence of ion induced cluster formation in the energy measurements of direct recoiling clusters from ion bombarded surface [6]. The work presented here is an extension of our earlier study. Keeping the irradiating ion constant i.e., Xe+, we have varied the ion energy and dose with the irradiations done at grazing incidence. From the energy spectra the recoil cross sections $d\sigma_r/d\Omega$ for the emission of different regimes of clusters have been computed.

## B. Emission from Direct Recoil compared with collision cascade sputtering

The constituents of a surface recoiling in a binary collision with an incident ion are the primary knock-ons also known in radiation damage theory as the Direct Recoils-DRs carries a characteristic energy which is a function of target to projectile mass ratio (m2/m1) and the angle of recoil θ_r with respect to the projectile direction and the bombarding energy $E_0$. The energy of graphite atoms of mass $m_2$ recoiling at angle $\theta_r$ is given by Bohr [7] as $E_r = 4[(m_1 m_2)/(m_1 + m_2)^2] E_0 \cos\theta^2$; where $m_1$ and $E_0$ are the mass and energy of the projectile. The cross section for a particular recoil to occur for a combination of $(m_2/m_1)$, $\theta_r$ and the bombarding energy $E_0$ can be worked out from ion-solid collision theory and compared with experimentally obtained values. Sigmund [8] by utilizing the LSS theory [9] has approximated the differential recoil cross section $d\sigma(E_0, E_r)$ for small energy transfer $E_r$ ($E_r \ll E_0$) to the target atoms as $d\sigma(E_0, E_r) \propto \left(E_0^{-k} E_r^{-1-k}\right) dE_r$, where $k$ is a constant appropriately chosen for a given interaction. The presently reported $E_r$ measurements are performed at recoil angle $\theta_r = 87.8°$. Such a large recoil angle was chosen because $E_r$ diminishes for larger values of $\theta_r$ thereby considerably enhancing $d\sigma(E_0, E_r)$. The Direct Recoils being the primary events of the energetic projectile-target interaction subsequently initiate collision cascades in which the energy is shared with other neighbours in the solid. Most of the sputtering yield is due to ejections from solids upon interaction of these collision cascades with the surface. Sputtering yield theories [8,10] predict that the yield $S$ or the total flux of atoms sputtered in all directions and energies for unit flux of incident ions is directly proportional to the deposited energy $F_d$ and inversely to the surface binding energy $E_b$ of the target atoms as $S \propto F_d/E_b$. The density of energy deposition at the surface $F_d$ can be further estimated [8] by using the nuclear stopping cross section $S_n(E_0)$; $F_d \approx \nu \eta S_n(E_0) S_n(E_0)$ where $\nu = \nu(m_2/m_1)$ is a target to projectile mass dependent parameter and $\eta$ is the target density. Using the SRIM2000 code [11] the nuclear stopping cross section $S_n(E_0)$ is estimated between 80 and 240 eV/Å for 50-150 eV Xe+ incident on graphite.

A comparison of the two above mentioned mechanisms of producing ion-induced particle emission from solids clearly indicates that whereas the DRs have well defined energies, trajectories and points of origin, the sputtered particles have broad energy distribution peaked at $E_b/2$ [10] and originate from various trajectories and origins. The time scales of the two interactions are also widely different; individual DR events are over in $10^{-14}$-$10^{-15}$ s, while the cooling down of the heavy ion induced collision cascade can take up to $10^{-12}$ s. The two types of projectile-initiated collision events are distinct and therefore, the energy spectrum of DRs can positively discriminate against the low energy sputtered particles. In our experimental set up we detect the direct recoils while the sputtered particles are suppressed in the detection process. However, continuous erosion and modifications of the bombarded target surface due to sputtering with yields between 20-40 C atoms/Xe+ [12] has to be taken into consideration while analyzing DR energy spectra.

## C. Detection of Direct Recoils

Our detection of direct recoils depends on their being positively charged so that the available electrostatic, time-of-flight or momentum analysis could be utilized. The charge state of a $DR$ depends on factors like the type of chemical bonding of targets, formation of molecular orbitals during collision [13] and the state of sputtering surface e.g., adsorption, ion induced roughing etc.

Datz and Snoeks pioneering work [14] using an Ar+ ion beam in the 40-80 keV range on metallic targets used a momentum analyzer for the detection of DRs. Rabalais and co-workers have used the time-of-flight technique to look at direct recoils from low $E_0 (\leq 10 keV)$ ion bombarded graphite [15]. Eckstein and Mashkova [16] have utilized an electrostatic analyzer for the analysis of DRs from 5 keV noble gas ion bombarded graphite. All of the three techniques for the detection of charged recoils have their relative merits; whereas, time-of-flight method needs pulsed beams, the energy and momentum analyses normally use continuous beams. The electrostatic analyzer can detect recoiling clusters with $E_r$ up to tens of keV [6]. Momentum analysis of clusters is desirable to unambiguously characterize the q/m values but the required magnetic fields however, become unrealistically large for experimental arrangements like ours. For example, in case of $\theta_r = 80°$, a large magnet is needed with $B\rho = 4 \ [Tm]$, where B is the magnetic field in Tesla [T] and $\rho$ is the radius of curvature in meters [m] for resolving clusters like $C_{60}$. A comparison of the electrostatic versus the momentum analysis for the detection of DRs is discussed in ref. [17].

In this paper we are reporting a series of experiments where we have investigated mechanisms of production as well as monitoring of clusters recoiling from amorphous graphite surfaces under Xe+ ion bombardment by using electrostatic analysis. The important feature of these experiments is that carbon clusters have been generated by the heavy ions. The evolution of clusters and their subsequent fragmentation under continuing ion bombardment is revealed by detecting various clusters in the energy spectra of the DRs emitted as a result of subsequent collisions between ions and surface constituents.

## II. Experimental

The experimental set up is shown in figure (1). A $250 keV$ heavy ion facility has been used for the experiments. A $Xe^+$ ion beam with diameter $\leq 1 mm$ and energy between $50-150 keV$ can be delivered to a target at any desired angle $\alpha$ with the surface normal. The accelerator can deliver a few $\mu A$ collimated beam on the target. The pressure in the target chamber is better than $10^{-7} mbar$ maintained with a combination of ion and oil diffusion pumps. The target chamber is pumped with the ion pump and the liquid nitrogen trapped diffusion pumps provide vacuum in the pre-target chambers. During ion bombardment the background pressure rises to $5 \times 10^{-6} mbar$. In fig. (1) the beam and the recoil particles collimators with $\pm 0.1°$ divergence are shown with a $90°$ Electrostatic Energy Analyzer (EEA) and a Channel Electron Multiplier (CEM).

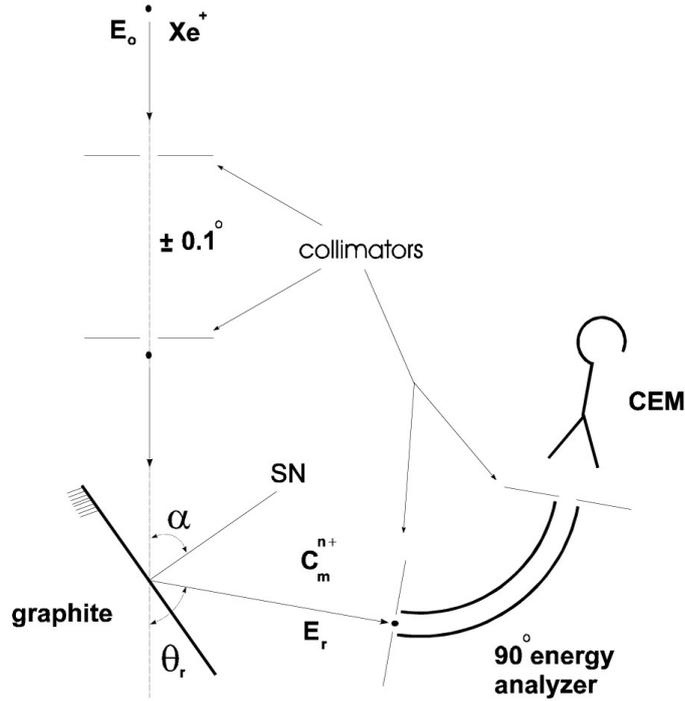

Figure (1). The experimental set-up where the beam as well as the recoil particles' collimators with $\pm 0.1°$ divergence are shown along with a 90° Electrostatic Energy Analyzer (EEA) and a Channel Electron Multiplier (CEM). A large angle $\alpha$ with the surface normal SN($\alpha = 80°$) is chosen for the present experiments.

Resolution of the electrostatic analyzer is $0.02$ with $0.8mm$ entrance aperture for the EEA. CEM with a typical gain of $10^7$ feeds the charged recoil data to a PC via a rate meter and a Hydra Data Acquisition unit. The condenser plate potential is increased in variable steps through a function generator (Philips PM 5138). The solid angle is fixed at $d\Omega = 6 \times 10^{-6}$ $st.rad$. Momentum analysis was also performed with a magnet with $B.\rho = 0.06$ $T.m$ and has been reported in ref. [17]. This analyzer is appropriate only for a recoil angle $\theta_r = 87.8°$. For smaller recoil angles much larger values of $B.\rho$ are required. In ref. [17] it was shown that the momentum analysis favors smaller masses and the heavier clusters are grouped at the end of the spectrum while the situation is exactly opposite in the case of recoil energy spectra. Therefore, in our experimental conditions and cluster identification requirements, energy analysis is preferred over the corresponding technique of momentum analysis.

The amorphous graphite targets are cut and polished from the glassy vitreous carbon discs provided by Le Carbone of UK. It has a density $\rho$=1.44 g.cm$^{-3}$ which is much below the normal graphite density of 2.26 g.cm$^{-3}$. The 10mm x 10mm samples are mechanically polished on various grit size SiC papers and the thickness of samples is brought to within 1mm $\pm$10%. The final finish provided by 1 $\mu$m paste on cloth. The sample is washed and vacuum degassed at 400-500 °C prior to irradiations. The XRD analysis of the samples before and after the irradiations does not show any evidence of crystallinity.

The ions are incident on the target at small angles with the surface i.e., $\alpha \geq 70°$. All our spectra were taken at $\alpha = 80°$.. In this case the shape of the irradiated region is a solid wedge with an angle of 10° between the ion path and the target surface. The adjacent sides are equal to the range $R$ of Xe+ in C at $E_0$=50-150 keV. Correspondingly the maximum depth from the surface of the irradiated region is $R\sin\alpha \approx$ 70-165 Å ($R$ is calculated from ref. [11]). We have $\mu A$ of collimated Xe+ beam incident on the amorphous C target. Using the extrapolated values of sputtering yields for $Xe^+ \rightarrow C$ from ref. [12] approximately 2-4 mono-layers are sputtered per second. Therefore, we have a set of competing processes that are initiated due to the energy deposition by Xe+ in an elliptical region whose length is R and the other dimension can be approximated by the transverse range straggling ~60-110 Å.

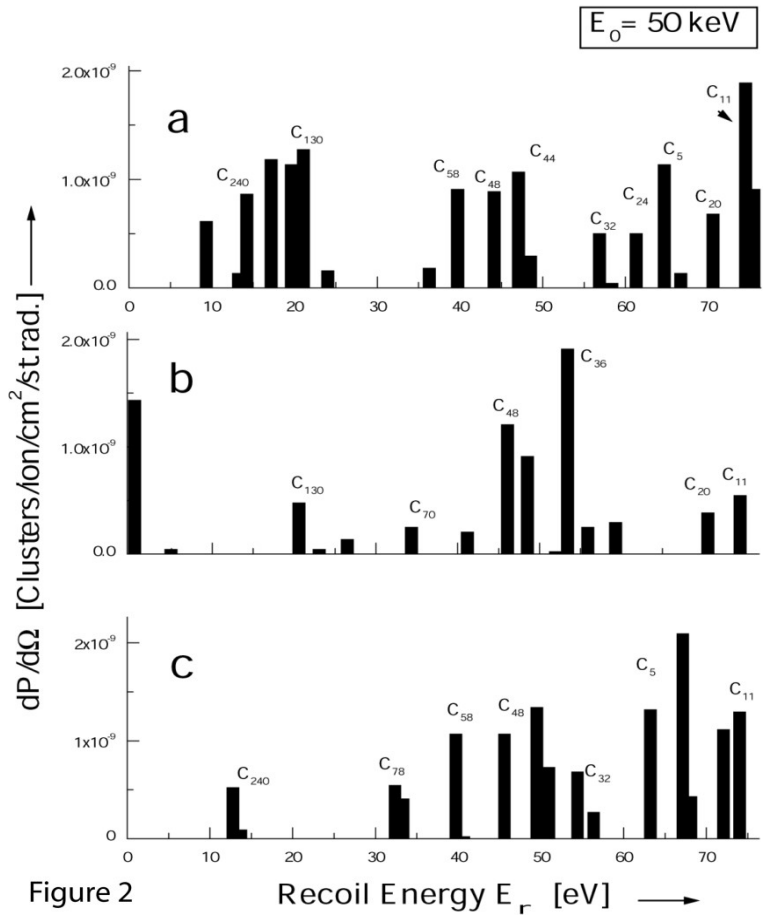

Figure 2

## III. Results

In figure (2) three direct recoil energy spectra are shown from 50 keV Xe+ bombardment. Fig. (2-a) indicates that pockets of clusters $C_m$ with $m \leq 36$ and $m \geq 106$ are predominant in this initial spectrum. The central group of clusters with 36≤*m*≤106 is not dominant. In figs. (2-b) and (2-c), we observe a significant decrease in the numbers of heavier $C_m$ (i.e. *m*≥106) by almost an order of magnitude while the lower *m* clusters gradually dominate.

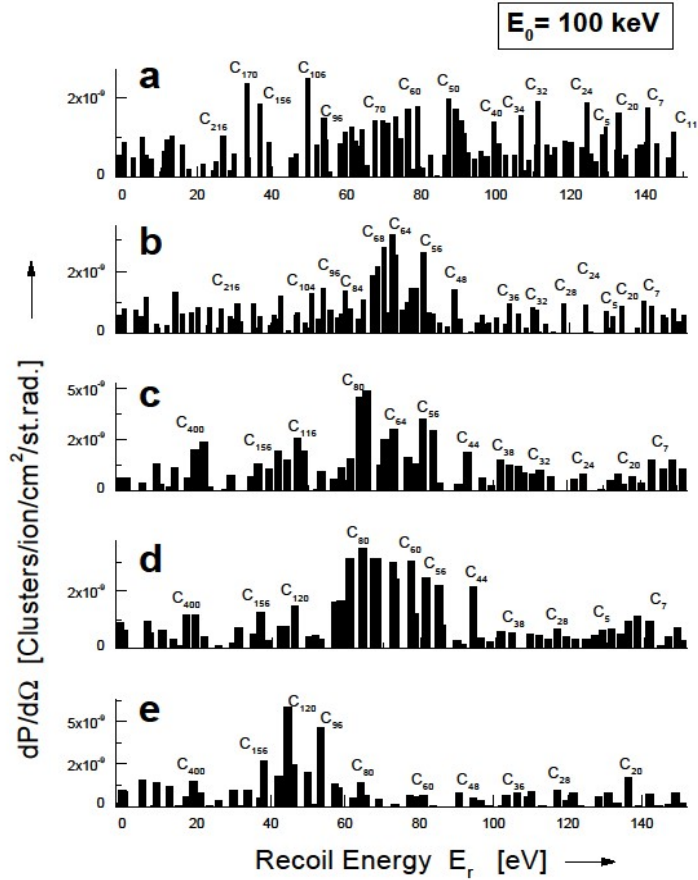

Figure (3). A set of five spectra each, labelled (a) to (e) are shown from 100 keV Xe+ bombardment. These have been obtained sequentially after intervals of 25 minutes each for a fluence of 2.2 x $10^{17}$cm$^{-2}$. In fig. (3-a) to (3-e), the lower order clusters with m≤36 do not increase their relative fraction of the respective spectra. On the other hand, the heavier clusters (m≥ 106) show a gradual increase in later spectra (3-d) to (3-e). The group with 36<m≤106 first increases in intensity from one spectrum to another, and subsequently decreases.

At $E_0$=100 keV, two sets of spectra are presented. The first ones (fig. 3) are obtained for a new target location with no prior ion implantation and the other ones (fig. 4) are taken with a heavily irradiated target. In both these figures a set of five spectra each, labelled (a) to (e) have been obtained one after the other consecutively within 25 minutes intervals corresponding to a fluence of 2x$10^{17}$ cm$^{-2}$ each. In fig. (3-a) to (3-e), the lower-order clusters with *m*≤36 are predominant as the bombardment continues and the cumulative dose increases with time at a flux of 1.46 x $10^{14}$ cm$^{-2}$.

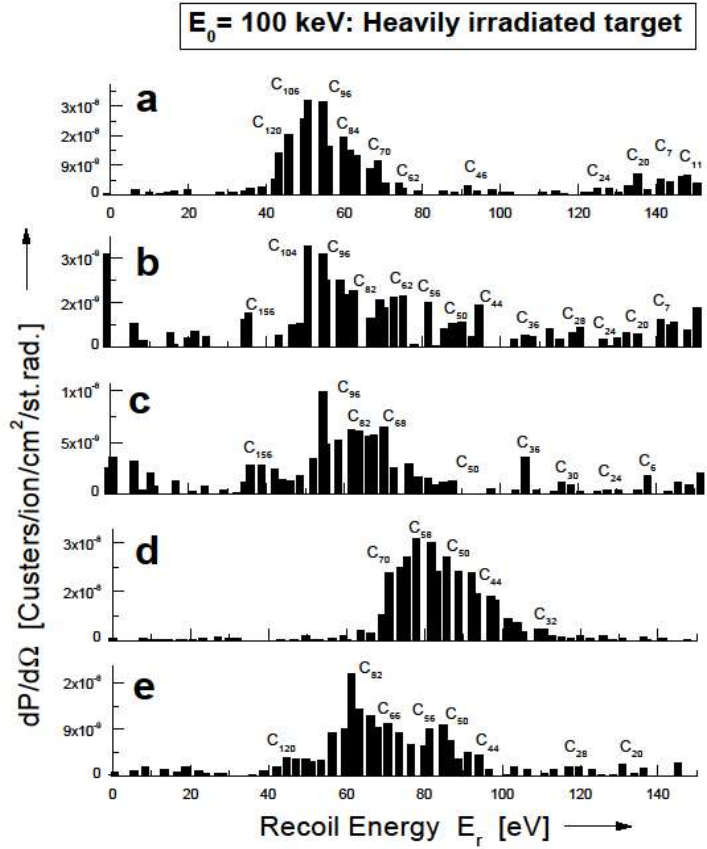

Figure (4). dP/dΩ versus $E_r$ spectra from a target heavily pre-irradiated for $10^4$ s with a 100 keV $Xe^+$ beam up to a cumulative fluence of ~$10^{18}$ cm-2. Maintaining a fluence of 2.2 x $10^{17}$ cm$^{-2}$ per spectrum we observe that in figs. (4-a) to (4-c) the clusters are centered around 36<m≤106. In figure (4-d) the clusters dominate around 32<m<72.

The heavier clusters (m>106) show a gradual increase in the later spectra. The group of clusters with 36≤$m$≤106 is slowly increasing as a function of the dose but the individual cluster contributions do not show significant variations. This indicates that clustering and fragmentation processes are both of the same order of magnitude.

Figure (4, a-c) is the set of direct recoil spectra from a target which has been continually irradiated for about $10^4$ s with 100 keV $Xe^+$ beam on the previous day. Figure (4) demonstrates the clustering and later fragmentation sequences under continuous ion irradiation. In fig. (4-a) to (4-c) the clusters in the mass range 36≤$m$≤106 are centered around $C_{96}$. It is evident from the figures (4-a) to (4-c) that for the heavier clusters, the intensity first decreases and then increases again as seen e.g. in (4-c) for $C_{82}$. Figure (4-d) peaks around $C_{70}$ and $C_{32}$, with an order of magnitude higher emission probability as compared with those in figure (4-b).

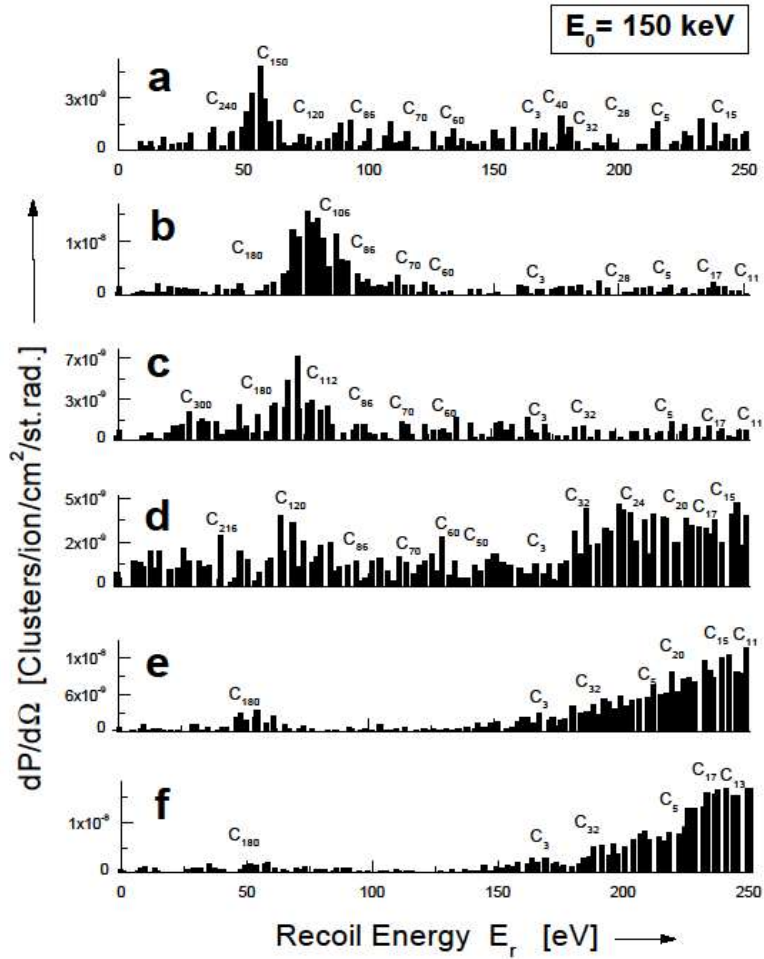

Figure (5). The set of six spectra from (5-a) to (5-f) is taken with $E_0$=150 keV at the constant fluence per spectrum of fluence of 8 x $10^{16}$ cm$^{-2}$. Figure (5-a) has a prominent peak around $C_{150}$ with other clusters being present in comparable numbers. The next two spectra show first a build up of clusters with a broad massive peak around $C_{106}$ in fig. (5-b) and fragmentation in the later spectrum (5-c). The next three spectra from (5-d) to (5-f) present the gradual build up of the lighter clusters i.e., 2<m≤36. Notice the change of scale in the last two spectra.

Figure (5) is obtained with $E_0$=150 keV. The set of six spectra from (5-a) to (5-f) show the gradual dominance of lighter clusters at the expense of the heavier ones. Starting with figure (5-a) we see that the spectrum has a prominent peak around $C_{150}$, with other clusters being present in comparable numbers. The next two spectra show first a build-up of clusters with a broad massive peak around $C_{106}$ in (5-b), and fragmentation in the next spectrum (5-c). This fragmentation is seen to spread out the main group of clusters towards the lighter ones. The next three spectra from (5-d) to (5-f) have shown the gradual build up of lighter clusters i.e., *m*≤36 dominating over the heavier ones.

## IV. Discussion

To summarize and present the data shown in figures (2) to (5) table 1 computes the recoil cross sections $d\sigma_r/d\Omega$ for $Xe^+ \rightarrow C$ at $E_0$=50, 100 and 150 keV. The recoil cross section is obtained from the data by using $d\sigma_r/d\Omega = dP/d\Omega(1/\rho\delta x)$; where $\rho$ is the target density and $\delta x$ the irradiated target area. The peak intensity from the plot of $dP/d\Omega$ vs. $E_r$ for a particular cluster is then converted into the recoil cross section $d\sigma_r/d\Omega$. The table deals exclusively with three cluster regimes m= 36, 36<m≤ 106 and m>106.

### A. The missing sputtered $C_1^+$ peak from the DR spectra

Keeping in view the large sputtering yield and the correspondingly large cross section for the emission of low energy (~$E_b$/2) $C_1$ particles one expects a massive $C_1^+$ peak. This peak has neither been seen as a major contributor to the energy spectra in figs. (2) to (5) presented in this paper nor in the experiments reported earlier [6,17]. To resolve this apparent anomaly, we have conducted a special series of experiments [18] employing the twin techniques of the ion induced photon emission spectroscopy and the mass analysis of the sputtered +ve and -ve C ions. Our results clearly show that whereas, the characteristic photon emission spectra identifies $C_1^+$ along with the excited neutral monatomic carbon $C_1^0$, the excited diatomic carbon $C_2^0$ peaks; the mass spectrum does not have $C_1^+$, $C_{21}^+$ peaks. In fact, only the negative carbon species $C_1^-$, $C_2^-$, $C_3^-$,... have been detected. On the basis of these new results, we have interpreted the DR spectra without the elusive $C_1^+$ (sputtered) monatomic carbon in the positively charged species.

### B. Energetic Xe+ induced topographical changes in the amorphous C surface

Changes in the surface topography of the heavily irradiated graphite are directly related to the rather high sputtering yield (20-40 atoms/ion) for energetic Xe+. The grazing ion incidence less than 10° between the ion beam and the surface, further reduces the effective penetration depth transverse to the surface thereby restricting the damage to ~ 100 Å beneath the surface. With $\rho$=1.44 gcm-3 our sample is much less tightly packed than the normal graphite with $\rho$ = 2.26 gcm-3. We would like to point to yet another major difference between the sputtered and direct recoiled species from such a dynamical surface re-building process. Typical values of the sputtering yield per unit solid angle $dS/d\Omega$ for 100 keV $Xe^+ \rightarrow C$ is $\sim 10^6$ $C_1/Xe^+/st.rad$. Comparison of this value can be done with $dP/d\Omega \sim 10^{-9} C_m/Xe^+/st.rad$ from the data presented in e.g., fig. (3). The 15 orders of magnitude difference in the rates of the two ion induced target particle emission mechanisms clearly shows that while sputtering is the main agent for the topographical changes, the direct recoils carry the instantaneous surface constitution information.

## C. Evolution of clusters in the heavy ion bombarded graphite

By using experimental sputtering yields S for $Xe^+ \rightarrow C$ [12] we can see that S~20-40 C atoms/ion for grazing incidence. Such a high sputtering yield combined with the $I_{ion}$=3$\mu$A on a target of 0.128 cm$^2$ implies a target surface erosion rate of ~ 2-4 layers s$^{-1}$. High energy irradiation leads to a competition between clustering and fragmentation as we have seen in the figures (2) to (5). The DR yield is therefore associated with recoils generated by the incident ions from a highly irradiated target surface that has gone through various sequences of bond re-arrangements. A particular DR event is representative of the changes that have preceded the specific collision. Typically a 100 keV Xe$^+$ ion has a range of ~660Å and takes ≈10$^{-15}$ s to deposit its energy before coming to a stop. Spacing between successive primary recoils is ~ 50 Å and the primary recoil energy distribution $\propto E_r^{-3/}$ [10]. For $Xe^+ \rightarrow C_1$ the carbon recoil energy $E_r(C_1)$ varies from ~$E_b$ at $\theta_r(max) \approx \pi/2$ to 15.4 keV for $\theta_r(min) \approx \pi/4$. This corresponds to the range of projectile scattering angle $\varphi$ between $\varphi(min) \approx 0.1°$ and $\varphi(max) \equiv sin^{-1}(m_2/m_2) = 5.24$. A heavy projectile's path in lighter targets is almost straight and the primary knock-ons are generated within cones with half angles =$\theta_r$. Since the majority of these ion-target atom collisions favour low energy recoils, the forward moving recoils in these cones have half angles $\theta_r$>$\pi/4$ in the case of $Xe^+ \rightarrow C_1$. The cone density along the track follows from the primary recoils energy distribution according to $cos^3\theta_r$. Thus the lower $E_r$ and high $\theta_r$ primaries are wrapped around the ion path with a constant linear density. The energy spectrum of these primaries is further convoluted in the subsequent $C_1 \rightarrow C_1$ collisions with the characteristic scattering and recoil angle $\pi/2$ and cascading of collision events with a linear recoil density $\propto E_r^{-2}$.

As a consequence of these forward cones' formation and the generation of isotropically expanding collision cascades the bombarded region is in a state of disorder due to the irreversible atomic movements and rearrangements of local $C - C$ bonds. These structure modifying dynamics of the atoms still leaves the irradiated regons in an amorphous state and the disorder persists. The high energy density with predominant nuclear energy loss is responsible for the dynamics of the surface reconstruction processes. We would like to refer to Klaumunzer and Schumacher's pioneering work [19] on ion beam induced deformation in metallic glass $Pd_{80}Si_{20}$ and on a range of other amorphous materials including semiconductors, metallic and insulating glasses [20]. The important difference in their work and our is the energy of projectile. They have used $100-350 MeV$ projectiles while ours have three orders of magnitude lesser energy. The energy deposition mechanism at kinetic energies $\geq 1 Mev/amu$ is due to electronic energy loss therefore, it has been argued [20] that the intense electronic excitations provoke atomic rearrangements and are responsible for the drastic strucural deformations. The ion induced physical and chemical changes in our case on the other hand, can be related mostly to the nuclear stopping powers $S_n$ (145-160 eV/Å) while the $S_e$ is in the range 27-41 eV/Å [11] for $Xe^+ \to C_1$ at $50, 100$ and $150 keV$. We can conclude that the process described here is the consequence of nuclear energy transfer. The similarities between the two irradiation induced surface effects lie in the resulting surface features. In both the cases surfaces at macroscopic level develop wave like configuration which is stable under further irradiation. This is a unique effect seen only in irradiated amrphous materials that undergo a transition from randomly ordered solid surfaces to surface wave patterns that still have a random arrangement of atoms on microscopic level.

## V. Conclusions

Our DRS results on intensely irradiated amorphous graphite indicate the dynamic surface rebuilding processes. The experimental data presented in fig.(2) to (5) and the computed values of direct recoil cross sections $d\sigma_r/d\Omega$ in table 1 identify the formation and emission of clusters under ion bombardment of graphite as a function of ion energy and dose. The energetics of $C-C$ bond formation in a highly disordered region has been proposed to be dependent on the nuclear energy dissipation processes. Within the deformed regions the reorganization of the energetic carbon atoms in excited or ionized states may lead to cluster formation $\sum_m C_m \to C_m$ along with their subsequent fragmentation $C_m \to C_m - C_2$. Direct recoils that originate in binary collisions between incident ions and the surface constituents carry with them the information that characterizes the dynamics of cluster formation and fragmentation within the reconstructed surface.

# References

[1]. D. Fink, M. Muller, L. T. Chadderton. P. H. Cannington, R. Elliman and D. C. McDonald, Nucl. Instrum. Meth. in Phys. Res. **B 32** [2] D. Fink, K. Ibel, P. Goppelt, J. P. Biersack, L. Wang and M. Behar, Nucl. Instrum. Meth. in Phys. Res. **B 46**, 342 (1990).

[3] G. Brinkmalm, D. Borofsky, P. Demirev, D. Fanyo, P. Hakansson, R. E. Johnson, C. T. Reimann and B. U. R. Sundqvist, Chem. Phys. Lett. **191**, 345 (1992).

[4] R. M. Papaleo, A. Hallen, P. Demirev, G. Brinkmalm, J. Eriksson, P. Hakansson and B.U.R. Sundqvist, Nucl. Instrum. Meth. in Phys. Res. **B 91**, 677 (1994).

[5] L. T. Chadderton, D. Fink, Y. Gamaly, H. Moeckel, L. Wang, H. Omichi and Y. Hosoi, Nucl. Instrum. Meth. in Phys. Res. **B 91**, 71 (1994).

[6] M. N. Akhtar, B. Ahmad and S. Ahmad, Phys. Lett. **A 234,** 367 (1997).

[7] N. Bohr, Dan. Vid. Selsk Mat. Fys. Medd. **18** (1948) No. 8.

[8] P. Sigmund, Topics in Applied Physics, Vol. 47, Sputtering by Particle Bombardment I, Ed. R. Behrisch, ( Springer-Verlag, Berlin, 1981) pp 9-71.

[9] J. Lindhard, M. Scharff and H. E. Schiott, Mat. Fys. Medd. Dan. Vid. Selsk., **33**, (1963) No. 14.

[10] M. W. Thompson, Defects and Radiation Damage in Metals (Cambridge University Press, Cambridge, 1969) chap. 4 & 5.

[11] J. F. Ziegler, J. P. Biersack and U. Littmark, The Stopping and Range of Ions in Solids, Vol. 1 (Pergamon Press, New York, 1985).

[12] H. H. Andersen and H. L. Bay, Topics in Applied Physics, Vol. 47, Sputtering by Particle Bombardment I, Ed. R. Behrisch, (Springer-Verlag, Berlin, 1981) pp 145-218.

[13] W. J. Lichten, Phys. Chem. **84,** 2102 (1980).

[14] S. Datz and C. Snoek, Phys. Rev. **134**, 347 (1964).

[15] J. W. Rabalais and Jie-Nan Chen, J. Chem. Phys. **85,** 3615 (1986).

[16] W. Eckstein and E.S. Mashkova, Nucl. Instrum. Meth. in Phys. Res. **B 62,** 438 (1992).

[17] A Qayyum, B. Ahmad, M.N. Akhtar, S. Ahmad, Eur. Phys. J. **D 3**, 26 (1998).

[18] A Qayyum, M. N. Akhtar, T. Riffat, S. Ahmad, Appl. Phys. Lett. **75**, 4100 (1999).

[19] S. Klaumunzer, G. Scumacher, Phys. Rev. Lett. **21**, 1987 (1983).

[20] Ming-dong Hou, S. Klaumunzer, G. Shumacher, Phys. Rev. **B 41**, 1144 (1992).